\documentclass[runningheads]{llncs}

\usepackage{graphicx}
\begin{document}
\title{sim2real: Cardiac MR Image Simulation-to-Real Translation via Unsupervised GANs}
\makeatother

\author{Sina Amirrajab\inst{1} \and Yasmina Al Khalil\inst{1}
\and
Cristian Lorenz\inst{2}
\and
Jürgen Weese\inst{2}
\and
Josien Pluim\inst{1}
\and
Marcel Breeuwer\inst{1,3}
}

 %
\authorrunning{S. Amirrajab et al.}
\titlerunning{sim2real: Cardiac MR Image Simulation-to-Real Translation}
\institute{Eindhoven University of Technology, Eindhoven, The Netherlands
\and
Philips Research Laboratories, Hamburg, Germany\and
{Philips Healthcare, MR R\&D - Clinical Science, Best, The Netherlands}
}
\maketitle              
\begin{abstract}
There has been considerable interest in the MR physics-based simulation of a database of virtual cardiac MR images for the development of deep-learning analysis networks. However, the employment of such a database is limited or shows suboptimal performance due to the realism gap, missing textures, and the simplified appearance of simulated images. In this work we 1) provide image simulation on virtual XCAT subjects with varying anatomies, and 2) propose sim2real translation network to improve image realism. Our usability experiments suggest that sim2real data exhibits a good potential to augment training data and boost the performance of a segmentation algorithm.

\keywords{Cardiac Image Simulation  \and Style transfer \and Image Synthesis.}
\end{abstract}
\section{Introduction}
A cohort of virtual cardiac magnetic resonance images (CMR) can be simulated to aid the development and adaptation of data-hungry deep-learning (DL) based medical image analysis methods. Recent studies have shown the effectiveness of image simulation in the context of training a DL model for CMR image segmentation \cite{xanthis2021simulator,SASHIMI}. Although such models provide accurate anatomical information, their performance is still suboptimal as a result of the realism gap, missing texture and simplistic appearance of the simulated images. This holds especially for models trained completely on simulated images and evaluated on real ones. Generative adversarial networks (GANs) \cite{goodfellow}, on the other hand, promise to synthesize realistic examples, as demonstrated by applications for multi-modal medical image translation \cite{bauer2021generation,jin2019deep,CT2MR_Heart}. However, GAN-generated images may not necessarily represent plausible anatomy.

The purpose of the current research is to reconcile the two worlds of simulation and synthesis, as defined in \cite{Frangi}, and take advantage of recent developments in the field of computer vision to reduce the realism gap between simulated and real data using GANs for unpaired/unsupervised style transfer. The contributions are two-fold: 1) Physics-based simulation of cardiac MR images on a population of XCAT subjects 2) GANs-based image-to-image translation for style (texture) transfer from real images. The framework is named sim2real translation.

\section{Material and Method}

The 4D XCAT phantom \cite{XCAT} is utilized as the basis of the anatomical model for creating virtual subjects by carefully adjusting available parameters for altering the geometry of the human anatomy. We employ our in-house CMR image simulation framework based on the analytical Bloch equations to generate varying image contrast on the labels of the XCAT virtual subjects \cite{SASHIMI}.

\begin{figure}[!ht]
    \centering
    \includegraphics[width=0.90\textwidth]{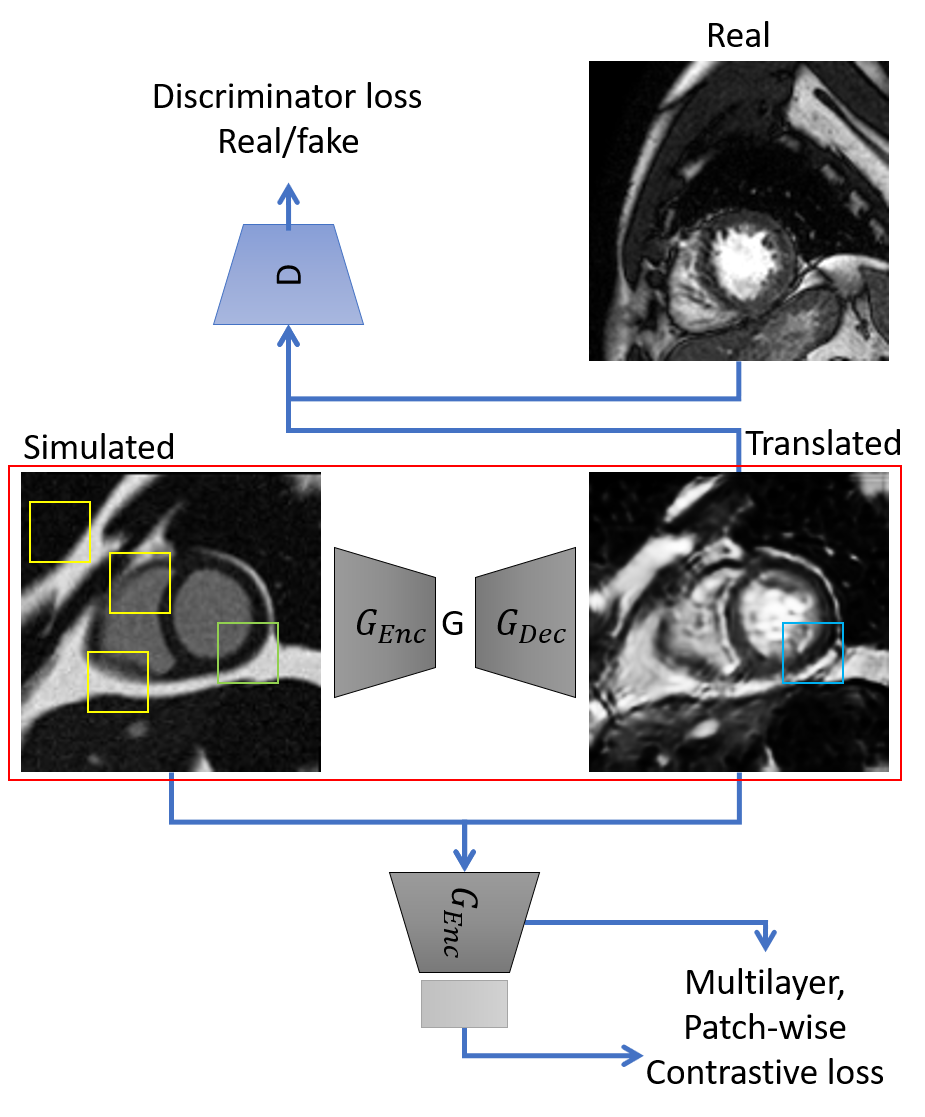}
    \caption{simulation-to-real (sim2real) translation using contrastive learning for unpaired translation model proposed in [14]. The style translation in achieved by the discriminator loss, while the content of anatomical information in the simulated data is preserved by the multilayer, patch-wise contrastive loss}
    \label{fig:cut_model}
\end{figure}

An unsupervised GAN model based on contrastive learning, known as CUT \cite{park2020contrastive}, is used for the task of unpaired translation between the real and the simulated images to transfer the realistic style (texture) from real images to simulated ones while preserving the anatomical information (content). Contrastive learning encourages encoded features of two patches from the same location in the real and translated images to be similar yet different to other patches. Compared to other unpaired translation frameworks such as CycleGAN \cite{CycleGAN}, CUT is a one-sided network with a much lighter generator architecture hence requiring few data for training. The content of the simulated image is preserved through a multilayer patch-wise contrastive loss added to the adversarial loss, as shown in Figure \ref{fig:cut_model}.

The M\&Ms challenge data \cite{M&Ms} are used as the source of real cardiac MR images. To explores the effects of multi-vendor data, we utilize a subset of the data consisting of 150 subjects with a mix of healthy controls and patients with a variety of heart conditions scanned using Siemens (Vendor A) and Philips (Vendor B). We extract four mid-ventricular slices at end diastolic (ED) and end systolic (ES) phases for each subject. All subjects are resized, centre cropped to the size of 128 x 126, and normalized to the range of [0, 1]. The same pre-processing is applied on the simulated images, despite the fact we use the available ground truth labels of the simulated data to find a bounding box around the heart and crop accordingly instead of centre cropping.

Two identical sim2real models are trained using the data from vendor A and vendor B (sim2real A, and sim2real B) to investigate the network’s ability to transfer vendor-specific appearance images on simulated ones. We calculate the widely-used Fréchet Inception Distance (FID) score \cite{heusel2017gans} between feature vectors calculated for real and translated images to evaluate the similarity between the simulated database and its respective real data, before and after translation. 

Additionally, we evaluate the usefulness of our sim2real data in aiding a DL segmentation model for the task of cardiac segmentation. We utilize a nnUNet \cite{nnunet}, trained to segment the left ventricle (LV), right ventricle (RV), and the left ventricular myocardium (MYO). First, we train a model using 150 sim2real images with the style of vendors A and B and compare it to a model trained on 150 real images.  We additionally train a model with a mixed set of real and sim2real data to observe the applicability of generated data for data augmentation.

\begin{figure}[!ht]
    \centering
    \includegraphics[width=0.93\textwidth]{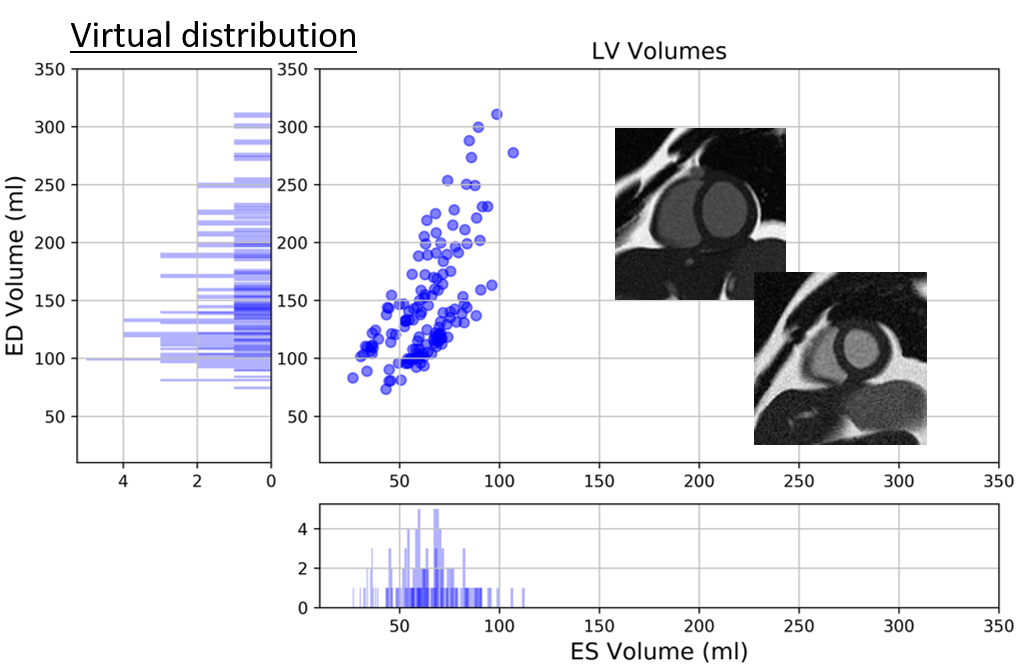}
    \caption{End-diastolic and end-systolic volume of left ventricle for simulated virtual subjects and two examples of cardiac MR image simulation.}
    \label{fig:ED_ES_volumes}
\end{figure}

\begin{figure}[!ht]
    \centering
    \includegraphics[width=0.93\textwidth]{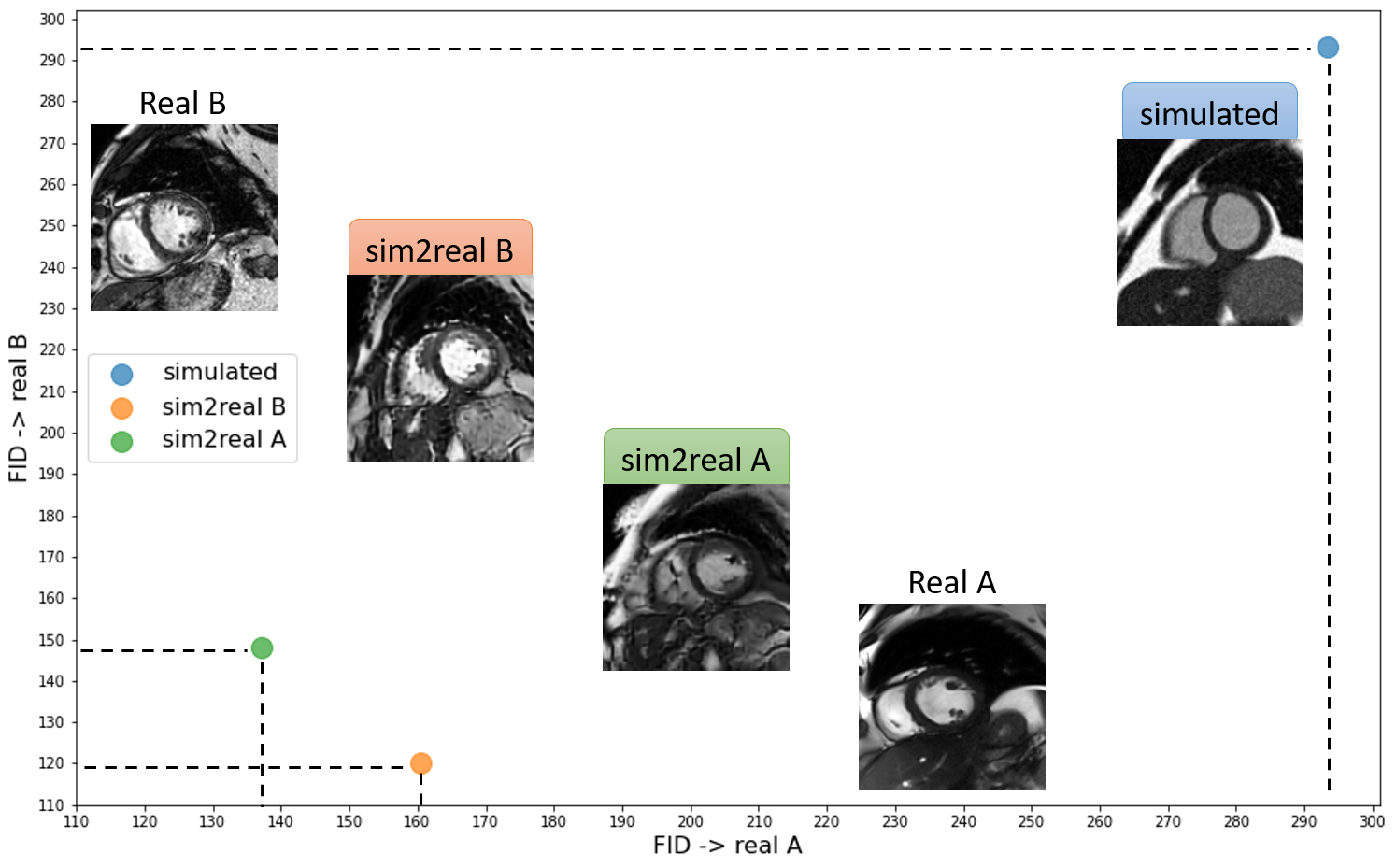}
    \caption{Fréchet inception distance (FID) score, which lower value means more similarity, between simulated and real data from vendor A (FID $->$ real A), simulated and real data from vendor B (FID $->$ real B), style transfer from vendor A on simulated denoted as sim2real A, and style transfer from vendor B on simulated denoted as sim2real B. One real example for each domain is also shown.}
    \label{fig:FID_score}
\end{figure}

\section{Results}
Two examples of simulated images and statistics of the XCAT virtual subjects’ distribution in terms of left ventricular volumes are depicted in Figure \ref{fig:ED_ES_volumes}.

The FID score is computed between the simulated data, sim2real A data, sim2real B data and the data from vendor A and vendor B. The lower value for the FID score suggests more realistically generated images and thus higher similar feature statistics to real database. The results are shown in Figure \ref{fig:FID_score}. As expected, the original simulated data has a high FID score on both real A and real B data. Generally, the sim2real model substantially reduces the FID between the simulated data and real images, indicating improvement in image realism. Moreover, the vendor-specific imaging features are captured by the network and transferred to the simulated images. One real example from each vendor and each sim2real translation is shown for visual comparison. 

\begin{table}[!ht]
    \centering
    \resizebox{0.98\linewidth}{!}{%
    \begin{tabular}{cc||cc|cc|cc|cc|cc|cc}
     \multicolumn{2}{c||}{Training} &\multicolumn{12}{c}{Testing}  \\\hline
     \multicolumn{2}{c||}{} &\multicolumn{6}{c}{Vendor A} &\multicolumn{6}{|c}{Vendor B}  \\\hline
     \multicolumn{2}{c||}{} &\multicolumn{2}{c}{LV} &\multicolumn{2}{c}{RV} &\multicolumn{2}{c|}{MYO}&\multicolumn{2}{c}{LV} &\multicolumn{2}{c}{RV} &\multicolumn{2}{c}{MYO}  \\
     Real & Simulated & Dice & HD & Dice & HD & Dice & HD & Dice & HD & Dice & HD & Dice & HD  \\\hline
     N/A & N=160 & 0.887 & 9.25 & 0.851 & 12.45 & 0.801 & 14.72 & 0.871 & 10.38 & 0.861 & 11.21 & 0.831 & 12.11  \\
     N=160 & N/A & 0.901 & 8.19 & 0.878 & 9.35 & 0.863 & 9.88 & 0.893 & 9.31 & 0.872 & 10.67 & 0.849 & 9.76  \\
     N=160 & N=160 & 0.915 & 7.85 & 0.882 & 10.21 & 0.872 & 12.32 & 0.911 & 7.28 & 0.874 & 10.85 & 0.851 & 10.21  \\\hline
    \end{tabular}
    }
    
    \caption{Segmentation performance of 2D nnUNet models trained with only simulated data (row 1), only real data (row 2) and a mix of real and simulated data (row 3), where N indicates the total number of images used for training. All models are evaluated on the unseen test set from vendors A and B in terms of the average Dice score and Hausdorff Distance (HD) per three cardiac tissues. Results outlined in red indicate the best performing model per tissue.}
    
    \label{tab:seg_results}
\end{table}

The segmentation performance of three different models can be observed in Table \ref{tab:seg_results}, presenting the evaluation of all models on a separate test set from the M\&Ms challenge. The results suggest that the model trained with sim2real images already adapts well to real data, exhibiting a slight drop in performance compared to the model trained with real data. Additionally, we observe that augmenting the training with sim2real data has a positive impact on segmentation accuracy (Dice score), particularly for the LV. 

\section{Discussion and Conclusion}
In this work, we created a database of virtual cardiac MR images simulated on the XCAT anatomical phantom and investigated the effectiveness of an unsupervised GAN for the task of simulation-to-real translation, named sim2real. We attempted to reduce the realism gap between the simplified image simulation and complex realistic image textures. Our sim2real model could learn the vendor-specific imaging features and map them onto the simulated images, resulting in reduction of the FID scores which can be attributed to more similarity between the simulated and real databases. Our usability experiments suggested that sim2real data exhibits a good potential to augment real training data, particularly in scenarios where data is scarce. 

\bibliographystyle{splncs04}
\bibliography{bib}

\begin{thebibliography}{10}
\providecommand{\url}[1]{\texttt{#1}}
\providecommand{\urlprefix}{URL }
\providecommand{\doi}[1]{https://doi.org/#1}

\bibitem{bauer2021generation}
Bauer, D.F., Russ, T., Waldkirch, B.I., T{\"o}nnes, C., Segars, W.P., Schad,
  L.R., Z{\"o}llner, F.G., Golla, A.K.: Generation of annotated multimodal
  ground truth datasets for abdominal medical image registration. International
  journal of computer assisted radiology and surgery  \textbf{16}(8),
  1277--1285 (2021)

\bibitem{M&Ms}
Campello, V.M., Gkontra, P., Izquierdo, C., Martin-Isla, C., Sojoudi, A., Full,
  P.M., Maier-Hein, K., Zhang, Y., He, Z., Ma, J., et~al.: Multi-centre,
  multi-vendor and multi-disease cardiac segmentation: the m\&ms challenge.
  IEEE Transactions on Medical Imaging  \textbf{40}(12),  3543--3554 (2021)

\bibitem{CT2MR_Heart}
Chartsias, A., Joyce, T., Dharmakumar, R., Tsaftaris, S.A.: Adversarial image
  synthesis for unpaired multi-modal cardiac data. In: International workshop
  on simulation and synthesis in medical imaging. pp. 3--13. Springer (2017)

\bibitem{Frangi}
Frangi, A.F., Tsaftaris, S.A., Prince, J.L.: Simulation and synthesis in
  medical imaging. IEEE transactions on medical imaging  \textbf{37}(3),
  673--679 (2018)

\bibitem{goodfellow}
Goodfellow, I.J., Pouget-Abadie, J., Mirza, M., Xu, B., Warde-Farley, D.,
  Ozair, S., Courville, A., Bengio, Y.: Generative adversarial networks. arXiv
  preprint arXiv:1406.2661  (2014)

\bibitem{heusel2017gans}
Heusel, M., Ramsauer, H., Unterthiner, T., Nessler, B., Hochreiter, S.: Gans
  trained by a two time-scale update rule converge to a local nash equilibrium.
  Advances in neural information processing systems  \textbf{30} (2017)

\bibitem{nnunet}
Isensee, F., Jaeger, P.F., Kohl, S.A., Petersen, J., Maier-Hein, K.H.: nnu-net:
  a self-configuring method for deep learning-based biomedical image
  segmentation. Nature methods  \textbf{18}(2),  203--211 (2021)

\bibitem{jin2019deep}
Jin, C.B., Kim, H., Liu, M., Jung, W., Joo, S., Park, E., Ahn, Y.S., Han, I.H.,
  Lee, J.I., Cui, X.: Deep ct to mr synthesis using paired and unpaired data.
  Sensors  \textbf{19}(10), ~2361 (2019)

\bibitem{SASHIMI}
Khalil, Y.A., Amirrajab, S., Lorenz, C., Weese, J., Breeuwer, M.: Heterogeneous
  virtual population of simulated cmr images for improving the generalization
  of cardiac segmentation algorithms. In: International Workshop on Simulation
  and Synthesis in Medical Imaging. pp. 68--79. Springer (2020)

\bibitem{park2020contrastive}
Park, T., Efros, A.A., Zhang, R., Zhu, J.Y.: Contrastive learning for unpaired
  image-to-image translation. In: European conference on computer vision. pp.
  319--345. Springer (2020)

\bibitem{XCAT}
Segars, W.P., Sturgeon, G., Mendonca, S., Grimes, J., Tsui, B.M.: 4d xcat
  phantom for multimodality imaging research. Medical physics  \textbf{37}(9),
  4902--4915 (2010)

\bibitem{xanthis2021simulator}
Xanthis, C.G., Filos, D., Haris, K., Aletras, A.H.: Simulator-generated
  training datasets as an alternative to using patient data for machine
  learning: an example in myocardial segmentation with mri. Computer Methods
  and Programs in Biomedicine  \textbf{198},  105817 (2021)

\bibitem{CycleGAN}
Zhu, J.Y., Park, T., Isola, P., Efros, A.A.: Unpaired image-to-image
  translation using cycle-consistent adversarial networks. In: Proceedings of
  the IEEE international conference on computer vision. pp. 2223--2232 (2017)

\end{thebibliography}
\end{document}